\documentclass[paper,twocolumn]{jpsj3}
\usepackage{txfonts}
\usepackage{color}

\title{Fragmentation Statistics of Food Diced and Crushed Using a Food Mixer}

\author{Naoki Kobayashi\thanks{kobayashi.naoki75@nihon-u.ac.jp} and Hitoshi Shibayama}
\inst{College of Industrial Technology, Nihon University, 2-11-1 Shin-ei, Narashino, Chiba 275-8576, Japan} 

\abst{
The fragment-size distributions of raw carrot diced or crushed using a food mixer are studied experimentally.
For the 5-mm-square raw carrot, the normal distribution shows a characteristic feature of food fragmentation statistics. 
This simple result indicates that most random errors contribute to fragment-size fluctuation. 
On the other hand, for the crushed raw carrot, the cumulative fragment size distribution follows the power law where the exponent $\alpha \simeq 1.62 > 1$.
Furthermore, considering the cumulative fragment-size distribution as a function of length for comparison with geomaterials, such as fault rocks, 
 the exponent $D \simeq 3.64$.
Previous studies have shown that the power-law distribution observed in sequential fragmentation tends to have a large exponent value.
As our experiment is also based on sequential fragmentation, the obtained large values of exponents $\alpha$ and $D$ are consistent
 with those obtained in previous studies on sequential fragmentation.
On the basis of previous studies and our observations, we discuss the effect of the preferential fragmentation of particles
 as large as the mixer blades.
We also discuss the existence of a lower limit beyond which further fragmentation is difficult, resulting in a power-law distribution
 tendency for raw carrot crushed with a food mixer. 
}

\begin{document}
\maketitle

\section{Introduction}
Fragmentation ubiquitously occurs in many natural phenomena. It is highly complicated and it is difficult to understand its dynamics.
Investigation of fragment-size distribution may lead to the comprehension of the fragmentation process. 
The fragment-size distribution resulting from various types of fragmentation has attracted the interest of physicists \cite{Herrmann, A06}.
In particular, the universality of impact fragmentation is of interest because its statistics follow the power-law distribution,
 and the scaling exponent depends on the shape and dimensionality of an object\cite{OPB93, MB96}.
Fragment-size distribution has been extensively studied \cite{GB61, Klacka92, IM92, Hayakawa96, AHT00, KSH03, KA03, WKHK04, WCKH08}.

Besides single impact fragmentation, sequential fragmentation has attracted the interest of researchers in recent years.
For example, in both fault rocks and volcanic fall deposits, the size of fragments was determined on the basis of the sequential brittle fragmentation of constituent particles.
Previous studies have experimentally and numerically confirmed that in many cases, the fragment-size distribution
 follows the power-law distribution \cite{Turcotte, SKB87, KJ98, DMP12, KVLM13, JMERCSE16, JR17, KPVM18}.

However, the fragment-size distribution may not necessarily exhibit power-law behavior.
For example, sequential fragmentation follows not only the power law but also the stretched-exponential \cite{Brown89, BW95}
 and lognormal distributions \cite{Kolmogorov41}.
In particular, the lognormal distribution was observed when the impact energy applied to the object was small.
Impact fragmentation of glass rods and plates \cite{IM92, KSH03} showed that
the fragment-size distribution changed from a lognormal distribution to a power-law distribution depending on the impact energy.

Similarly, in human mastication, which can be regarded as sequential fragmentation where the impact energy for a single event is not very large,
 the fragment-size distribution of the food bolus follows a lognormal distribution \cite{KKSM06, KKS10}.
Mastication is an in-mouth process in which food is broken, ground, or crushed by the teeth before swallowing and digestion \cite{Bourne, Lucas}.
The chewed food fragments are assembled into a bolus by a complicated movement of the palate and tongue immediately before swallowing \cite{H04}.
From the viewpoint of physics, bolus formation is considered as random packing.
The physical properties of the bolus are affected by the packing fraction of the constituent particles \cite{KY18}.
Recently, the physical properties of the bolus have been studied to find ways for people
 with reduced masticatory function to swallow food easily \cite{Szczesniak02, PGHLDMW11, SKM11,MYKIN19}.
These previous studies mainly applied stress-strain tests to investigate the rheology of a food bolus.
On the other hand, there are only a few studies on the relationship between the distribution of fragment size and the physical properties of a food bolus.
Kohyama {\it et al.} showed that finely cut food, which is often used in nursing facilities,
 is not always an easy-to-eat food form as shown by the masticatory efficiency measured by electromyography.\cite{KNYYHS07}.
In addition, Kitade and coworkers experimentally investigated the relationship between the distribution of fragments after chewing and
 the values of texture properties such as cohesiveness of cubic agar gels of 3.5 and 15 mm sizes and agar gel crushed with a meat grinder as samples \cite{KKKM12, KKM13}.
The size distribution of the crushed gel follows a lognormal distribution, and there is a large variation in size. 
From the results of apparatus measurements, the cohesiveness of the crushed gel was high at any chewing frequency, whereas these of 3.5 and 15 mm agar gels
 were high only chewing frequencies.
The results indicate that the cohesiveness of a food bolus with varied fragment size is higher than that of a food bolus with
 a uniform size, which makes it possible to swallow safely. 
Furthermore, the change in cohesiveness with respect to the number of chews was slower for the 3.5 mm agar gel than for the 15 mm agar gel,
 and the number of food fragments tended to less likely increase.
This result suggests that there is a size at which food is not easily fragmentized by chewing.

In this study, we experimentally investigated the sequential fragmentation of raw carrot cut into 5 mm squares or crushed using a food mixer.
Both these prechopped carrots are offered in homes for the elderly.
We show that their fragments follow a normal distribution and power-law distribution with a specific exponent value.
Finally, we discuss a plausible fragmentation process for raw carrot crushed using a food mixer with reference to previous studies.

\section{Experimental Procedure}
Solid foods can be categorized into four groups: spongelike foods, gels, raw fruits and vegetables, and dry crispy foods \cite{KSHH04}.
We selected raw carrot from the raw fruit and vegetable group for our experiments.
In this study, two different shapes of raw carrot were studied.
In the first experiment, raw carrots were placed on a 5-mm-wide grid paper and cut with a kitchen knife, obtaining approximately 5 g of 5-mm-square samples. 
These samples were then spreadout on transparencies without overlap. Snapshots of the samples were captured using a digital camera ($\alpha 7$II, SONY, Tokyo) and digitized at a suitable resolution of $6000 \times 4000$ pixels.
In the next experiment, raw carrots were cut into 7 $-$ 9 g cylinders (diameter and height; 4 and 0.5 cm, respectively).  
The samples were then crushed in a food mixer (Iwatani, IFM-C20G, Tokyo) for approximately 1 and 5 s.
Crushed raw carrot fragments weighing 5 g were randomly taken and manually spreadout on transparencies so that they do not overlap.
Then, as in the first experiment, snapshots of the fragments were captured at the same resolution of $6000 \times 4000$ pixels.
Since it is difficult to precisely separate and spread very small fragments on transparencies, we ignore fragments smaller than 0.01 $\mathrm{cm}^2$
 when we examine fragment-size distributions below.

\section{Results and Discussion}
\subsection{Fragment-size distribution}
\noindent{\it 5-mm-Square Raw Carrot}.
\begin{figure}
\centering
\includegraphics[width=0.9\linewidth]{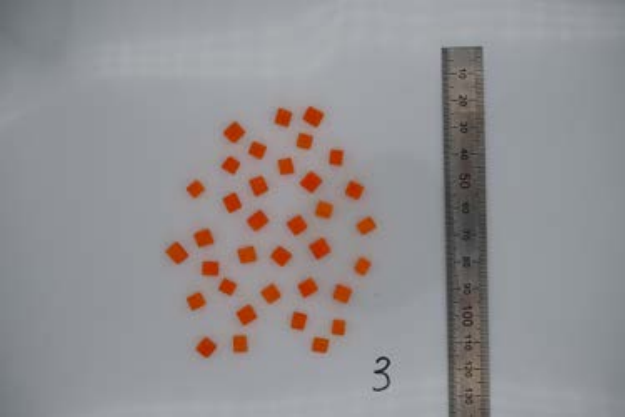}
\caption{Typical snapshot of 5-mm-square raw carrot sample fragments.}
\label{fig:Fig1}
\end{figure}
Figure 1 shows a typical snapshot of the 5-mm-square carrot fragments.
It can be observed that the surface area of each fragment is almost uniform.
Although 10 measurements were taken independently, the total number of fragments was 292, and the average number of fragments in each measurement was approximately 30.
The average surface area of the 10 integrated data was 0.383 ${\rm cm}^{2}$, and the standard deviation was $0.095$.
When preparing 5-mm-square carrot pieces, the length of each side was cut to 5 mm with a kitchen knife;
hence, the fluctuation of the side length will follow a normal distribution.
As the surface area is approximately the product of the sides, 
the fluctuation of the surface area will also follow a normal distribution. 
Furthermore, the data were assimilated and sorted by size, from the largest (i.e., the cumulative number).
Figure 2 shows the cumulative number of fragment sizes of 5-mm-square raw carrots.
\begin{figure}
\centering
\includegraphics[width=0.9\linewidth]{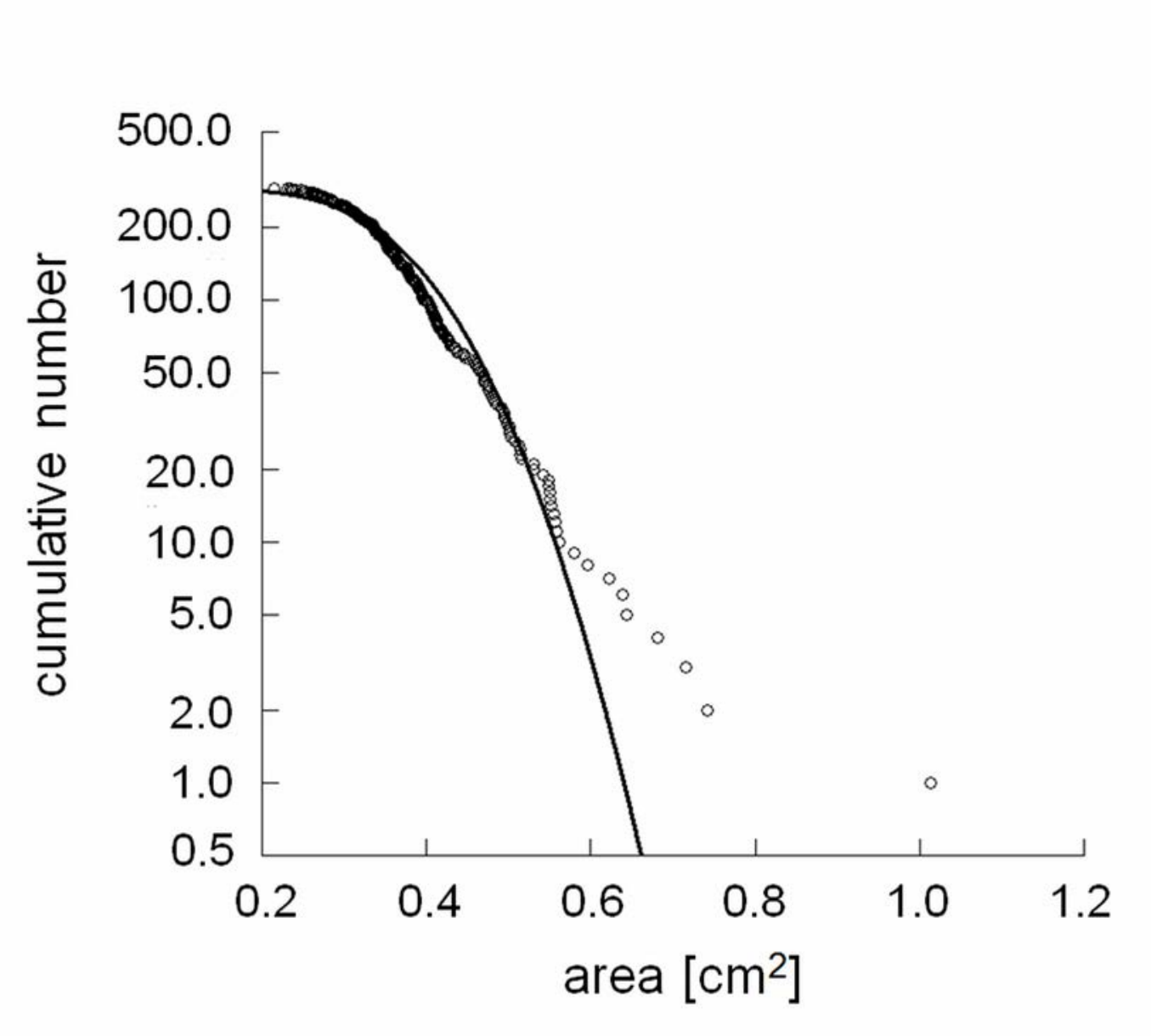}
\caption{Semilog plots of the cumulative number of 5-mm-square raw carrot fragments.
Additional data from 10 trials are shown.
The total number of fragments is 292.
The solid curve indicates the cumulative normal distribution for $N_{\rm{T}}=292, \bar{s}=0.383$, and $\sigma = 0.095$.}
\label{fig:Fig2}
\end{figure}
The cumulative size distribution $N(s)$ can be expressed as
\begin{align}
N(s) = \int_{s}^{\infty}n(s')ds',
\end{align}
where $n(s)$ is the fragment-size distribution.
As shown in Fig. 2, the cumulative fragment-size distribution is approximated by a cumulative normal distribution given by
\begin{align}
N(s) = \frac{N_{\mathrm{T}}}{2} \left(1 - \mathrm{erf}\left(\frac{s - \bar{s}}{\sqrt{2}\sigma}\right)\right),
\end{align}
where $N_{\mathrm{T}}$, $\bar{s}$, and $\sigma$ are the total number of fragments, mean, and standard deviation, respectively.
$\mathrm{erf}(x)$ is an error function defined as
\begin{align*}
\mathrm{erf}(x) \equiv \frac{2}{\sqrt{\pi}}\int_{0}^{x}e^{-y^2}dy.
\end{align*}
Note that the values of parameters $\bar{S}$ and $\sigma$ used in this fitting were obtained by measurements.\\
\\
{\it Crushed Raw Carrot}.
\begin{figure}
\centering
\includegraphics[width=0.9\linewidth]{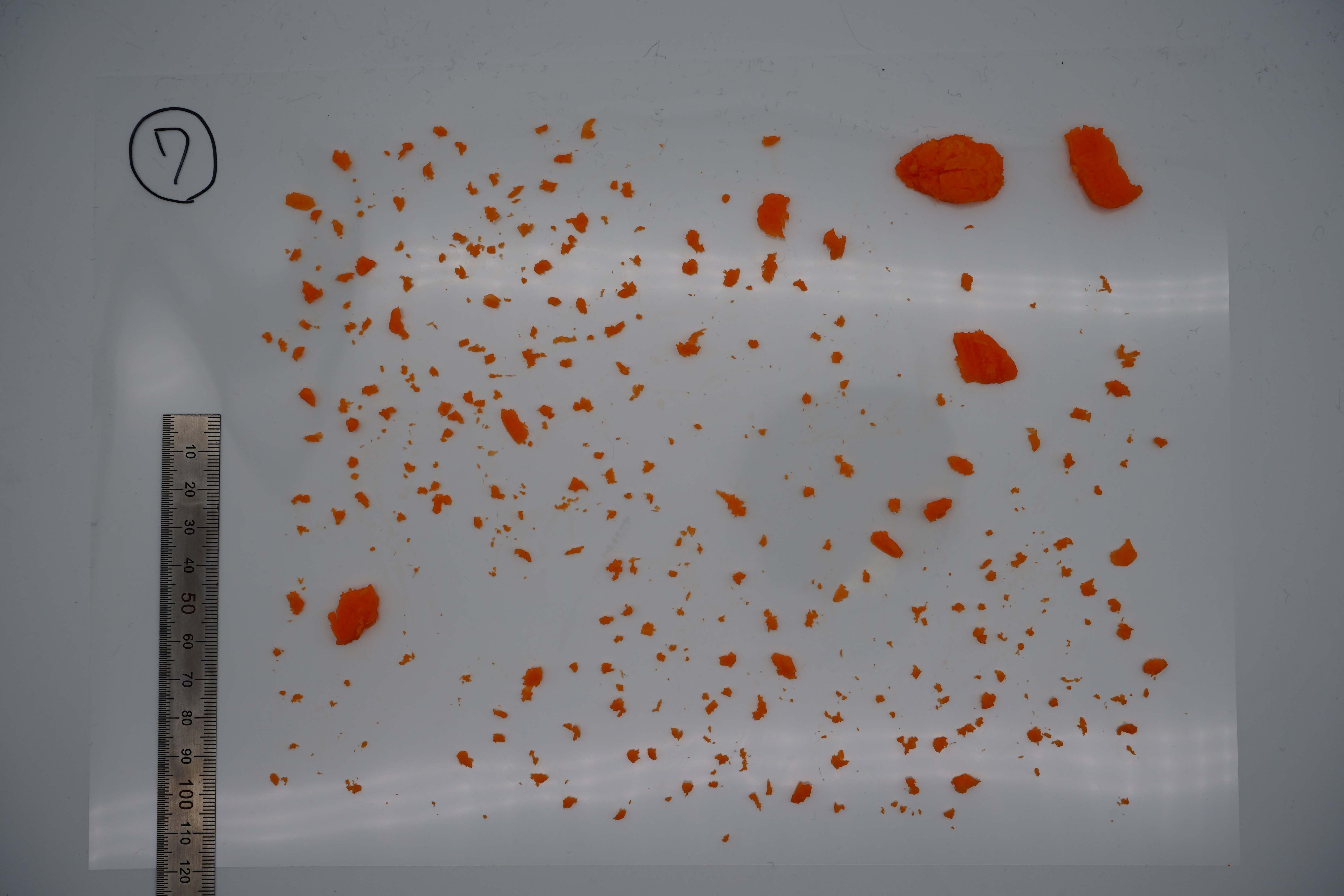}
\caption{Typical snapshot of the food fragment samples of raw carrot crushed using a food mixer.}
\label{fig:Fig3}
\end{figure}
Figure 3 shows a typical image of raw carrot crushed using a food mixer.
In contrast to Fig. 1, the size of food fragments is nonuniform.
The results of 10 integrated trials are described below.
The total number of food fragments was 3035, and the fragment size was distributed over a wide range from $10^{-2}$ $-$ 10 $\mathrm{cm}^2$.
Figure 4 shows the cumulative fragment size distribution, similar to Fig. 2. 
\begin{figure}
\centering
\includegraphics[width=0.9\linewidth]{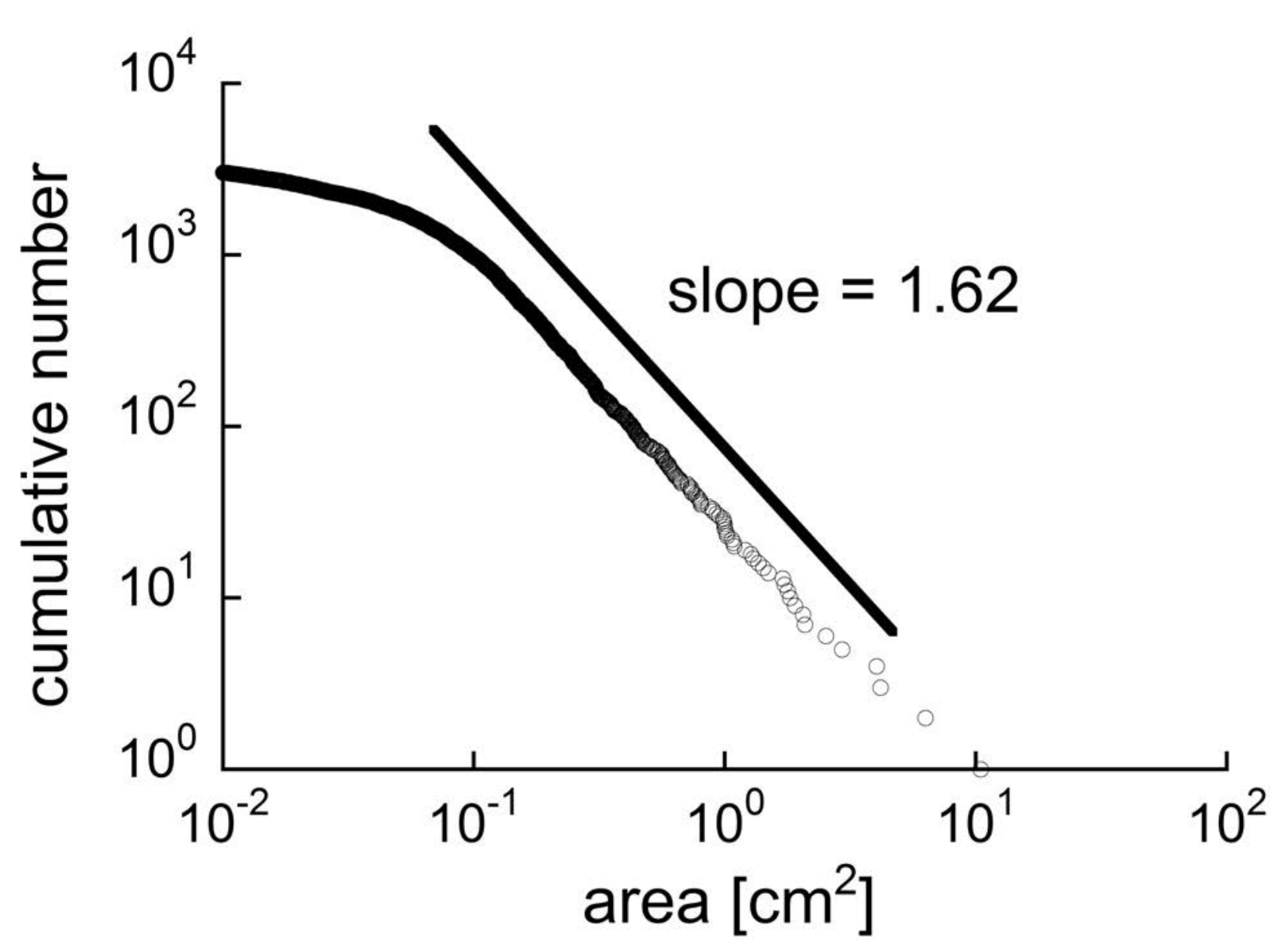}
\caption{Log--log plots of the cumulative number of crushed raw carrot fragments.
Additional data from 10 trials are shown.
The total number of fragments is 3035.
The least-squares fitted line gives $\alpha \simeq 1.62$.}
\label{fig:Fig4}
\end{figure}
The solid line in Fig. 4 is a fit of the power-law distribution given by
\begin{align}
N(s) \sim s^{-\alpha}.
\end{align}
The cumulative size distribution in Fig. 4 exhibits a clear power-law form in the $10^{-1}$ $-$ $10^{1}$ $\mathrm{cm}^2$ range,
 where the estimated value of exponent $\alpha$ is 1.62.
In this experiment, the surface area was used to evaluate the size of the fragments, rather than the mass,
 which was used in previous studies on fragmentation.
Ishii and Matsushita \cite{IM92} investigated the size distribution for the impact fragmentation of a glass rod,
 and they demonstrated that the length and mass of the fragments had almost the same value of the scaling exponent
 for the power-law distribution.
If the density is constant and there is no significant difference in thickness in three dimensions, the area and mass 
 can be assumed to be proportional.
In our experiment, although there are certain complications such as the shape of fragment crosssections, because the shape of
 individual fragments shown in Fig. 3 seem to be statistically isotropic, we can assume that this proportional relationship is valid.
Empirically, the exponent $\alpha$ of the cumulative size distribution is assumed not to exceed unity in accordance with Kor\v{c}ak's law \cite{Mandelbrot82, Feder88}.
Several previous studies on impact fragmentation also indicate a value that does not exceed unity \cite{GB61, OPB93, MB96, Hayakawa96, AHT00, KSH03, KA03, WKHK04, WCKH08}.
However, in our experiment, $\alpha > 1$.
In addition, the fragmentation of glass rods studied by Ishii and Matsushita \cite{IM92} follows a power-law distribution
 where the cumulative number of large fragments has an exponent value $\alpha > 1$.
In the experiment by Ishii-Matsushita, a glass rod was placed in a stainless-steel pipe.
A falling experiment was conducted, and sequential fragmentation of the glass rod in the stainless-steel pipe was clearly observed.
However, our result is consistent with the value obtained in previous studies on three-dimensional fragmentation under compression \cite{KVLM13, KPVM18},
which can be considered as sequential fragmentation because the fracture progresses in stages.

Previously Kun and coworkers \cite{KVLM13, KPVM18} numerically investigated the fragmentation of geomaterials.
Therefore, we compare our results with the scaling exponent commonly used in geomaterials.
Let us denote the cumulative number $N(r)$ as a function of $r$, with the square root of $s$.
If the scaling law $N(r) \sim r^{-D}$ is satisfied, then the value of exponent $D$ satisfies the relation $\alpha = \frac{D}{2}$
 with the exponent $\alpha$ of the fragment-area distribution \cite{Turcotte, Mandelbrot75}.
This relation yields an exponent value $D = 3.24$.
The large value of exponent $D$ obtained is consistent with those in previous studies on geomaterials where $D > 2$ \cite{Turcotte, SKB87, KJ98, DMP12, JMERCSE16, JR17}.
In most cases, such geomaterials are considered to be produced by sequential fragmentation.
Kaminski and Jaupart \cite{KJ98} mentioned that the large value of exponent $D$ for an explosive eruption is due to secondary fragmentation processes.
They experimentally showed that repeated fragmentation through particle--particle collision increases the value to $D > 3$.
Sammis {\it et al.} \cite{SKB87} suggested that the probability of fragmentation depends on the relative size of colliding fragments.
Such repeated size-dependent preferential fragmentation processes lead to a fractal distribution, i.e., the power-law distribution. \cite{Turcotte, Matsushita85, KA03}.
The food fragmentation in our experiment is considered as sequential fragmentation caused by not only the collisions between fragments
 but rather the collision between raw carrot and the blades of the food mixer stochastically.
Since the blades of the food mixer used in our experiment simply rotate in the same area, there are some fragments that are not fractured by the blades.
For example, the tail part of the size distribution in Fig. 4 shows that large fragments of about 1 -- 10 $\mathrm{cm}^2$ remained after processing
 without being fractured, suggesting that random selective fragmentation occured.
It is expected that the fragmentation process by the food mixer gives a power-law distribution with a large value of exponent.
However, from the size distribution shown in Fig. 4, it is not possible to confirm whether fragments with, for example, the size of the blades are selectively fractured.

Finally, let us examine the relationship between the crushing time of the raw carrot and the size distribution at that time.
Figure 5 shows the cumulative size distribution for food fragments crushed for approximately 1 and 5 $\mathrm{s}$.
\begin{figure}
\centering
\includegraphics[width=0.9\linewidth]{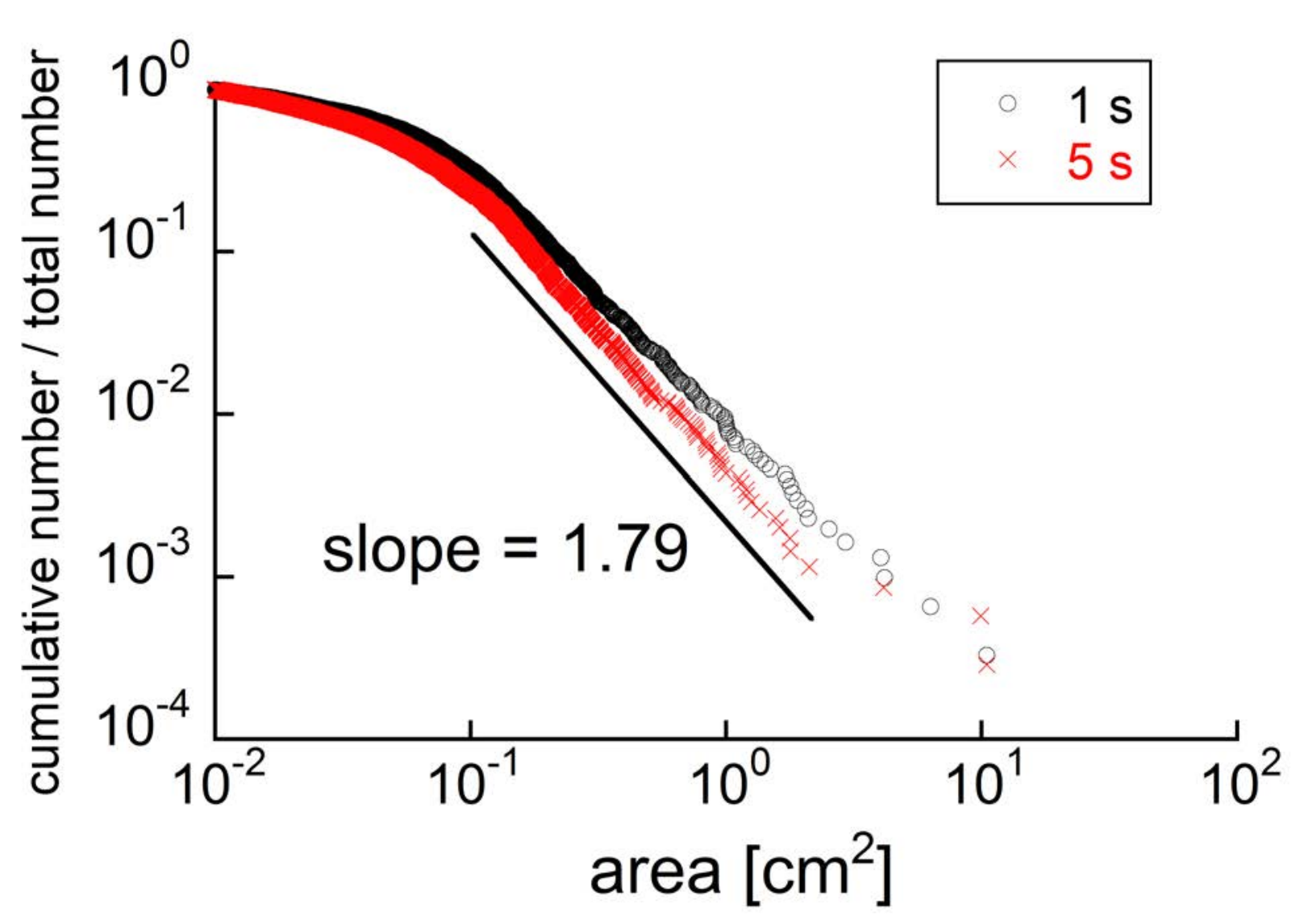}
\caption{Cumulative size distribution of fragments for crushing times of $1$ (circles) and $5$ (crosses) s.
Additional data from 10 trials are shown.
The total number of fragments for crushing time of $5$ $\mathrm{s}$ is 3463.
The least-squares fitted line gives $\alpha \simeq 1.79$.}
\label{fig:Fig5}
\end{figure}
Note that the vertical axis is normalized by the total number of fragments.
It can be seen that the cumulative distribution for the crushing time of 5 $\mathrm{s}$ in Fig. 5 shows a clear power-law form,
 similary to the crushing time of 1 $\mathrm{s}$.
The slope in Fig. 5 yields the value of exponent $\alpha$ for the crushing time of 5 $\mathrm{s}$, and we obtain $\alpha = 1.79$.

Matsushita and Sumida \cite{MS88} proposed simple stochastic models to better understand one-dimensional brittle fracture, such as those of glass rods \cite{IM92}.
Their models yield the fragment-size distribution close to the lognormal because their models have dynamics similar to these of Kolmogorov's cascade fracture \cite{Kolmogorov41}.
One of their models is a simple cascade model that leads to a power-law distribution from the lognormal by incorporating a lower threshold beyond
 which the fragments do not break when they become smaller than the threshold. 
The value of the exponent for their cascade model with the lower threshold depends on the total number of fragments, because of the accumulation effect
 due to having set the lower threshold.
In our experiment with crushed carrot, we can consider that the smaller fragments are softened by the water content of the raw carrot itself, rendering further fragmentation difficult.
In addition, our result shown in Fig. 5 suggests a positive correlation between the total number of fragments and the value of exponent $\alpha$,
 which is consistent with the result obtained with the models proposed by Matsushita and Sumida.
In the case of our experiment, we think that an effect similar to that in the Matsushita--Sumida models occurred.

\section{Concluding Remarks}
We studied the fragment-size distribution of raw carrot diced or crushed using a food mixer.
For the 5-mm-square raw carrot, we determined that the normal distribution shows a characteristic feature of food fragmentation statistics,
 as shown by the excellent fitting in Fig. 2.
This simple result shows that most random errors contribute to the fluctuation of fragment size.
On the other hand, for the crushed raw carrot, the cumulative fragment-size distribution followed the power law where the exponent $\alpha > 1$.
When the cumulative distribution is considered as a function of the length, the scaling also holds, and the scaling exponent is as large as $D \simeq 3.64$.
Previous studies have shown that the power-law distribution observed in sequential fragmentation tends to have a large exponent \cite{KJ98, SKB87, Turcotte, DMP12, KVLM13, JMERCSE16, JR17, KPVM18}.
As our result is also based on sequential fragmentation, it is suggested that sequential fragmentation under certain conditions,
 such as size-dependent preferential fragmentation \cite{SKB87}, leads to a power-law distribution with a large exponent.
In growing complex networks, it is known that the distribution of the number of nodes follows the power law with
 a large exponent value because of positive feedback action called preferential attachment \cite{KR01}.
The size distribution of the sequential fragmentation phenomenon follows a power-law distribution
 because a material of a certain size is selectively fractured \cite{SKB87, Turcotte, Matsushita85, KA03}, and there is {\it preferential} attachment of negative feedback as indicated by the reduction in size. 
In our experiment, the fragmentation process consists of sequential collisions between the raw carrot and the blades, so it is expected that
 fragments of similar sizes to the blade are selectively fractured.
However, in order to confirm the actual fragmentation process, other methods such as direct observation with a high-speed camera should be considered in addition
 to the analysis based on fragment-size distribution.

As described by Oddershede {\it et al.} \cite{OPB93} and Meibom and Balslev \cite{MB96} on regarding impact fragmentation, it has been shown that the initial shape
 of a sample to be fractured affects the scaling property.
On the other hand, the effect of initial conditions on sequential fragmentation has not been investigated in detail.
In our experiment, the initial shape of raw carrot is fixed to be cylindrical, but it will be necessary to study other shapes as well as the time dependence of the size distribution
 to clarify the universality in sequential fragmentation.

As mentioned in the introduction, the evaluation of physical properties, especially the cohesiveness, of the food bolus composed of fragments of various sizes
 is important for practical applications.
If we consider a food fragment as a wet granular material, the cohesiveness of the food bolus depends on the surface tension between wet fragments \cite{MN06}. 
However, it is difficult to define the cohesiveness of a bolus and to relate it experimentally to the information of wet fragments such as the fragment-size distribution. 
Therefore, mathematical modeling of bolus formation has been studied to understand it numerically \cite{PL97, KY18}.
In these models, bolus formation is reproduced assuming loose packing; but when considering the food oral processing \cite{Bourne, Lucas},
 it is expected to be reproduced assuming closed packing.
The relationship between the fragment-size distribution and the cohesiveness determined using the bolus formation model
 assuming the closed packing process should be examined in future studies. 

\section*{Acknowledgment}
N.K. was supported in part by a Grant-in-Aid for Scientific Research (No. 18K02248) from the Japan Society for the Promotion of Science.
One of the authors is grateful to M. Matsushita for many stimulating discussions.

\end{document}